# Intrinsic Spin Susceptibility and Pseudogap-like Behavior in Infinite-Layer LaNiO$_2$

D. Zhao[1], Y. B. Zhou[1], Y. Fu[3,4], L. Wang[3], X. F. Zhou[3], H. Cheng[3], J. Li[1], D. W. Song[1], S. J. Li[1], B. L. Kang[1], L. X. Zheng[1], L. P. Nie[1], Z. M. Wu[1], M. Shan[1], F. H. Yu[1], J. J. Ying[2], S. M. Wang[3], J. W. Mei[3,4,*], T. Wu[1,2,5,7,†] and X. H. Chen[1,2,5,6,7]

1. Hefei National Laboratory for Physical Sciences at the Microscale, University of Science and Technology of China, Hefei, Anhui 230026, China

2. CAS Key Laboratory of Strongly-coupled Quantum Matter Physics, Department of Physics, University of Science and Technology of China, Hefei, Anhui 230026, China

3. Shenzhen Institute for Quantum Science and Engineering, and Department of Physics, Southern University of Science and Technology, Shenzhen 518055, China

4. Shenzhen Key Laboratory of Advanced Quantum Functional Materials and Devices, Southern University of Science and Technology, Shenzhen 518055, China

5. CAS Center for Excellence in Superconducting Electronics (CENSE), Shanghai 200050, China

6. CAS Center for Excellence in Quantum Information and Quantum Physics, Hefei, Anhui 230026, China

7. Collaborative Innovation Center of Advanced Microstructures, Nanjing University, Nanjing 210093, China

*meijw@sustech.edu.cn
†wutao@ustc.edu.cn

**Abstract:**
**The recent discovery of superconductivity in doped infinite-layer nickelates has stimulated intensive interest, especially for similarities and differences compared to that in cuprate superconductors. In contrast to cuprates, although earlier magnetization measurement reveals a Curie-Weiss-like behavior in undoped infinite-layer nickelates, there is no magnetic ordering observed by elastic neutron scattering down to liquid helium temperature. Until now, the nature of the magnetic ground state in undoped infinite-layer nickelates was still elusive. Here, we perform a nuclear magnetic resonance (NMR) experiment through $^{139}$La nuclei to study the**

**intrinsic spin susceptibility of infinite-layer LaNiO$_2$. First, the signature for magnetic ordering or freezing is absent in the $^{139}$La NMR spectrum down to 0.24 K, which unambiguously confirms a paramagnetic ground state in LaNiO$_2$. Second, a pseudogap-like behavior instead of Curie-Weiss-like behavior is observed in both the temperature-dependent Knight shift and nuclear spin-lattice relaxation rate ($1/T_1$), which is widely observed in both underdoped cuprates and iron-based superconductors. Furthermore, the scaling behavior between the Knight shift and $1/T_1T$ has also been discussed. Finally, the present results imply a considerable exchange interaction in infinite-layer nickelates, which sets a strong constraint for the proposed theoretical models.**

The origin of high-temperature superconductivity in cuprates [1] remains a conundrum in condensed matter physics. Whether we can find cuprate analogs in connection with high-temperature superconductivity would be a very important step for this challenge [2]. Especially, the infinite-layer nickelates, which have the same crystalline structure and $3d^9$ electronic configuration as the infinite-layer cuprate CaCuO$_2$ [3], have already attracted attention from the high-$T_c$ community 20 years ago [4,5]. In fact, infinite-layer nickelates with the hard-to-stabilize Ni$^{1+}$ oxidation state could be indeed synthesized by chemical reduction from RENiO$_3$ (RE = La, Nd) [6,7,8]. After several failed attempts in LaNiO$_2$ thin film [9,10,11], superconductivity with $T_c$ up to 15 K was recently found in a Sr-doped infinite-layer NdNiO$_2$ thin film by Hwang's group [12]. This long-awaited breakthrough draws intensive interests back to verify its connection to cuprate superconductors and beyond [13-18].

One of the necessary prerequisites for understanding the connection of infinite-layer nickelates to cuprates is to figure out the electronic structure and magnetic properties. Although much theoretical progress has been immediately made to discuss the underlying physics of infinite-layer nickelates [13-18], experimental progress on both electronic structure and magnetic properties is still very limited in thin-film materials so far [19]. On the other hand, it is worth noting that, although superconductivity is still missing in the bulk infinite-layer nickelates [20,21], bulk materials instead of thin film might be a good choice to clarify the intrinsic magnetic properties in undoped infinite-layer nickelates. In fact, earlier magnetic susceptibility measurements have already revealed a Curie-Weiss-like behavior in infinite-layer LaNiO$_2$ [7] and NdNiO$_2$ [8]. However, bulk materials are always contaminated by magnetic impurities, which makes the intrinsic behavior of spin susceptibility still elusive. Nuclear magnetic resonance (NMR) is a fabulous tool for measuring intrinsic spin susceptibility in bulk materials. The magnetic impurity effect could be perfectly avoided in NMR measurement. Here, we perform a $^{139}$La NMR measurement on LaNiO$_2$ powder to study the intrinsic spin susceptibility.

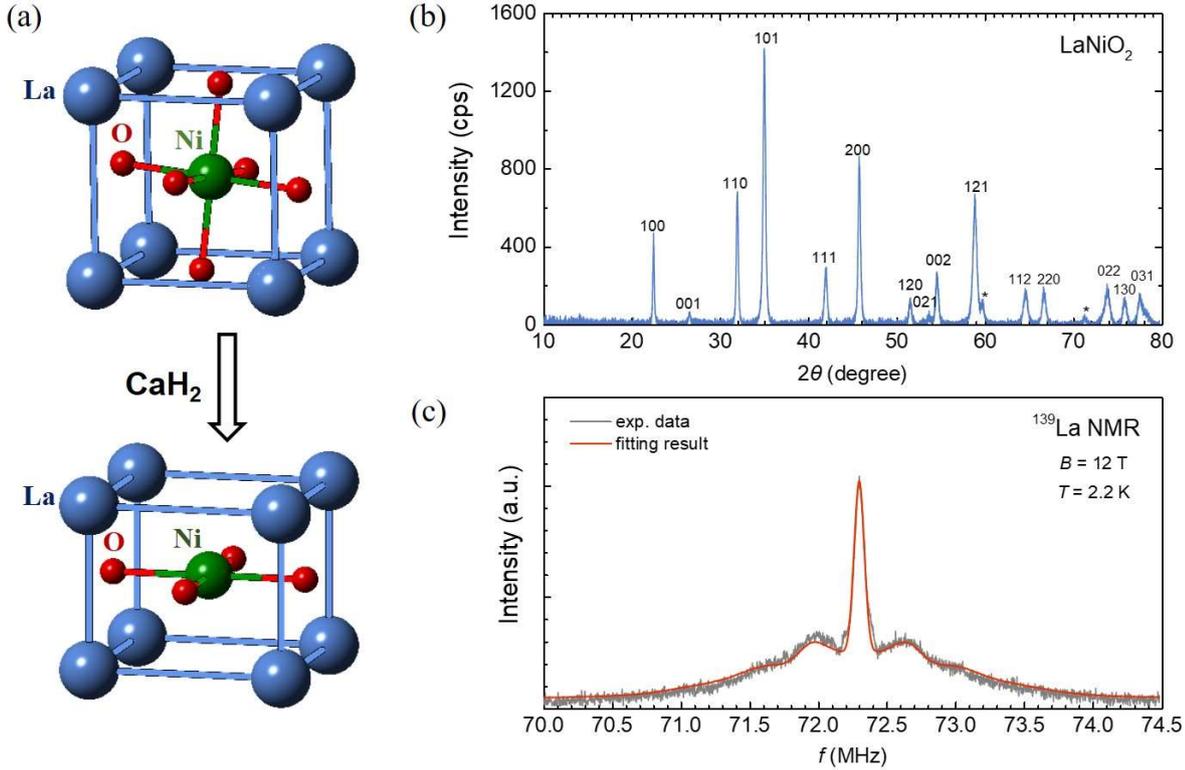

FIG. 1. Crystal structure, XRD result, and $^{139}$La NMR spectrum in LaNiO$_2$. (a) Sketch of structural change from LaNiO$_3$ to LaNiO$_2$ by topotactic reduction with metal hydrides CaH$_2$. (b) X-ray diffraction pattern of the LaNiO$_2$ powder sample. The main structural phase is proved to be LaNiO$_2$, and the asterisk indicates a small amount of impurity after topotactic reduction. (c) The full NMR spectra of $^{139}$La nuclei with spin number $I = 7/2$ and related fitting result (for details, see Sec. S4 in Supplemental Material[23]).

As shown in Fig. 1(a), the polycrystalline sample of infinite-layer LaNiO$_2$ is carefully synthesized by chemical reduction from LaNiO$_3$ powder, the same method as NdNiO$_2$ in the previous study [22] (see Sec. S1 in Supplemental Material for the detailed method[23]). The purity of both LaNiO$_3$ and LaNiO$_2$ has been checked by x-ray diffraction experiment as shown in Fig. S2 [23] and Fig. 1(b). In contrast to the pure LaNiO$_3$ precursor, there is a small amount of impurity phase in LaNiO$_2$, which should not affect the NMR result and analysis in the present work. The LaNiO$_2$ powder is directly sealed in a copper-wire-wound coil by epoxy adhesive to perform a NMR experiment. The external magnetic field is calibrated by $^{63}$Cu NMR with the same coil. In this work, we performed NMR measurement on the $^{139}$La nuclei whose nuclear spin number ($I$) is 7/2 and gyromagnetic ratio ($\gamma_N$) for bare nuclei is 6.0146 MHz/Tesla. As shown in Fig.1(c), the full NMR spectrum is quite broad and shows a symmetric multi-peak feature. By considering the quadrupole effect in the powder sample, we could perfectly fit the full NMR spectrum as shown in Fig.1(c). Compared to $^{139}$La NMR results of the LaNiO$_3$ precursor (as shown in Fig. S5 in Supplemental Material [23]), the extracted quadrupole frequency ($\nu_Q$) in LaNiO$_2$ is reduced to 0.8 from 1.46 MHz in LaNiO$_3$. In contrast, the relative quadrupole broadening ($\delta\nu_Q/\nu_Q$) in LaNiO$_2$ is increased to 20% from 9% in LaNiO$_3$. These results indicate that, after chemical reduction by CaH$_2$, the homogeneity of the chemical environment in LaNiO$_2$ becomes worse than that in LaNiO$_3$ precursor. In one previous NMR study, the quadrupole frequency of polycrystalline LaNiO$_3$ is reported to be only 1.1 MHz [35], which is smaller than the value of 1.46 MHz in the present study. Considering the possible change of oxygen content in LaNiO$_3$, such a difference suggests that the quadrupole frequency might be a sensitive probe for the oxygen content in LaNiO$_3$ and LaNiO$_2$. In fact, the oxygen content is a key factor to determine the electronic properties in the LaNiO$_{3-\delta}$ system [36]. Calibrating the oxygen content precisely would be very important to understand the relationship between oxygen content and electronic properties in the LaNiO$_{3-\delta}$ system. Here, the actual oxygen content in LaNiO$_3$ (LaNiO$_2$) is determined to be 2.93 (2.04) by using an electron probe x-ray micro-analyzer (for more details, see Sec. S1 in Supplemental Material [23]), which is close to ideal stoichiometry.

Next, we continue to study the temperature dependence of the NMR spectrum from 463 down to 0.24 K. Usually, when antiferromagnetic ordering emerges at low temperatures, an additional splitting or broadening effect in the NMR spectrum would be expected below the antiferromagnetic ordering temperature. As shown in Figs. 2(a) and 2(b), the central NMR spectrum does not show any additional splitting or broadening effect as the temperature decreases down to 0.24 K. This result indicates the absence of antiferromagnetic ordering in LaNiO$_2$, which is quite consistent with a previous elastic neutron scattering experiment at 1.7 K [7,8]. In addition, the temperature-dependent NMR linewidth also shows a weak and smooth temperature dependence [Fig.2(c)], suggesting the absence of significant spin freezing. In fact, the absence of spin freezing is also confirmed by the following nuclear spin-lattice relaxation measurement. We will return to this issue later.

Based on the temperature-dependent NMR spectrum in Fig. 2(a), we could extract the temperature-

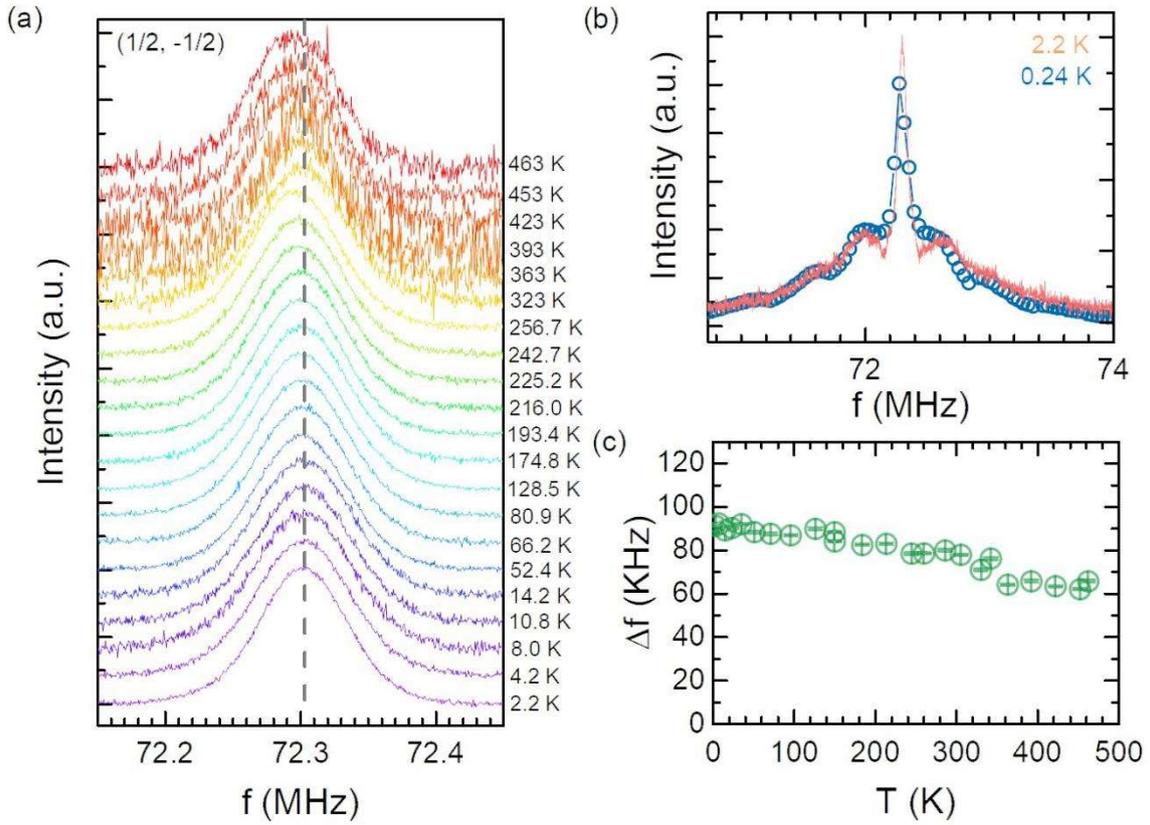

FIG. 2. Temperature-dependent $^{139}$La NMR spectrum in LaNiO$_2$. (a) Temperature-dependent central transition peak with the temperature range from 463 to 2.2 K. The value of external magnetic field is 12 T; (b) The comparison of full NMR spectrum between 2.2 and 0.24 K. No significant broadening effect has been observed; (c) Temperature-dependent linewidth of the central transition peak in (a). The linewidth is determined by a Gaussian fitting, and the corresponding error bar is the fitting error bar.

dependent averaged Knight shift ($K_{ave}$) (for more details, see Sec. S5 in Supplemental Material [23]). In principle, the magnetic part of the Knight shift can be divided into two main contributions. One main contribution is from spin shift ($K_s$), which is proportional to the uniform spin susceptibility [$\chi_s(q=0)$]. The other one is from orbital shift ($K_{orb}$), which is usually temperature independent. Usually, the temperature dependence of the magnetic Knight shift is consistent with the results from bulk susceptibility measurement. As shown in Fig. 3(b), by considering a negative hyperfine coupling at $^{139}$La sites, the temperature dependence of the Knight shift does not follow a Curie-Weiss-like behavior suggested by previous bulk susceptibility measurements [7,8,20]. Instead, the temperature-dependent

Knight shift decreases monotonically with decreasing temperature, which is quite similar to the pseudogap behavior observed in the underdoped cuprates [37]. Combined with a temperature-independent quadrupole contribution on the averaged Knight shift (for details, see Sec. S5 in Supplemental Material [23]), this result suggests that the Curie-Weiss-like behavior observed by bulk susceptibility measurement is extrinsic and should come from magnetic impurities. On the other hand, we have also tried to extract intrinsic magnetic susceptibility from $M$-$H$ curves at various temperatures by the Honda-Owen method (for more details, see Sec. S4 in Supplemental Material [23]). In contrast to previous bulk susceptibility measurements [7,8,20], the temperature dependence of extracted magnetic susceptibility from high-field limit is qualitatively consistent with the temperature dependence of the Knight shift by assuming a negative hyperfine coupling (see Fig. S4 in Supplemental Material [23]).

To further verify the pseudogap-like behavior in spin susceptibility, we further measure the temperature-dependent nuclear spin-lattice relaxation ($T_1$). In general, $1/T_1T$ is related to the dynamic spin susceptibility and has a general expression as $1/T_1T \sim \sum_q A_q^2 \cdot \chi_s''(q,\omega)/\omega$, where $A_q$ is the $q$-dependent hyperfine coupling tensor, $\chi_s''$ is the imaginary part of dynamic spin susceptibility and $\omega$ is the Larmor frequency. In the conventional Fermi-liquid scenario, $K_s$ is proportional to the density of states at the Fermi level [$N(E_F)$] and satisfies a so-called Korringa law, $1/T_1T \sim K_s^2$ [38]. As shown in Fig. 3(b), a similar pseudogap-like behavior is unambiguously confirmed in the temperature-dependent $1/T_1T$. Meanwhile, a similar temperature-dependent behavior in both the temperature-dependent Knight shift and $1/T_1T$ also suggests the absence of critical spin fluctuations. When the electronic system is approaching an antiferromagnetic (AFM) phase transition or quantum critical point (QCP), the dynamic spin susceptibility at the antiferromagnetic vector ($q_{AF}$) would be largely boosted due to critical spin fluctuations, which exhibits a exponential divergence or $1/(T - T_N)^\upsilon$ behavior for a two-dimensional AFM transition [39] and a specific power-law behavior for AFM QCP [40]. Then, the enhanced contribution at $q_{AF}$ to dynamic spin susceptibility would break the Korringa law. As shown in Fig. 3(a), a power-law behavior with $\alpha$ = 1.2 is observed in the temperature range from 100 down to 0.24 K, which is clearly not consistent with approaching a two-dimensional AFM transition [39]. If considering a possible AFM QCP, it always leads to a power-law behavior in temperature-dependent $1/T_1 \sim T^\alpha$ with power index $\alpha < 1$ [40]. However, our present result is also not consistent with any known antiferromagnetic critical behavior. Interestingly, a similar power-law behavior has been also observed in the temperature-dependent $1/T_1$ of $LaNiO_3$ [as shown in Fig. 3a] which also has no long-range antiferromagnetic ordering. In addition, significant spin freezing, which would lead to a pronounced peak-like behavior in the temperature-dependent $1/T_1$ around the freezing temperature [41], could be also excluded in our case. In fact, the fitting process of $T_1$ decay also suggests that the inhomogeneity of $T_1$ is getting worse below 100 K (see Fig. S7 in Supplemental Material [23]). A possibility due to quadrupole relaxation has been discussed in Supplemental Material and safely excluded (for more details, see Sec. S8 in Supplemental Material [23]). In our opinion, this might be related to inhomogeneous spin freezing at a very small scale, which should not be the intrinsic magnetic properties in LaNiO$_2$. Therefore, the present study suggests that the magnetic ground state of LaNiO$_2$ is a paramagnetic state far from antiferromagnetic phase transition or QCP. In Supplemental Material, the possible filtering effect due to a $q$-dependent hyperfine coupling tensor at the $^{139}$La site has also been calculated [23]. Strictly speaking, due to the filtering effect of the hyperfine coupling tensor at the $^{139}$La site, both G-type and stripe-type antiferromagnetism cannot be completely excluded in the present study [42], which needs further NMR measurement at different nuclear sites. However, if considering practical disorders in possible magnetic ordering structure, the NMR at $^{139}$La sites is still possible to exhibit some signatures even for G-type and stripe-type antiferromagnetism. This is the exact case for $^{89}$Y NMR to detect antiferromagnetism in YBa$_2$Cu$_3$O$_{6+x}$ [33,34]. As the temperature increases above 100 K, the temperature-dependent $1/T_1$ shows a distinct power-law behavior with $\alpha$ = 3, which leads to a rapid increase of $1/T_1T$ as the temperature increases [Fig. 3b]. Up to the highest measuring temperature, the Curie-Weiss-like behavior is still absent in temperature-dependent $1/T_1T$ as that in the Knight shift.

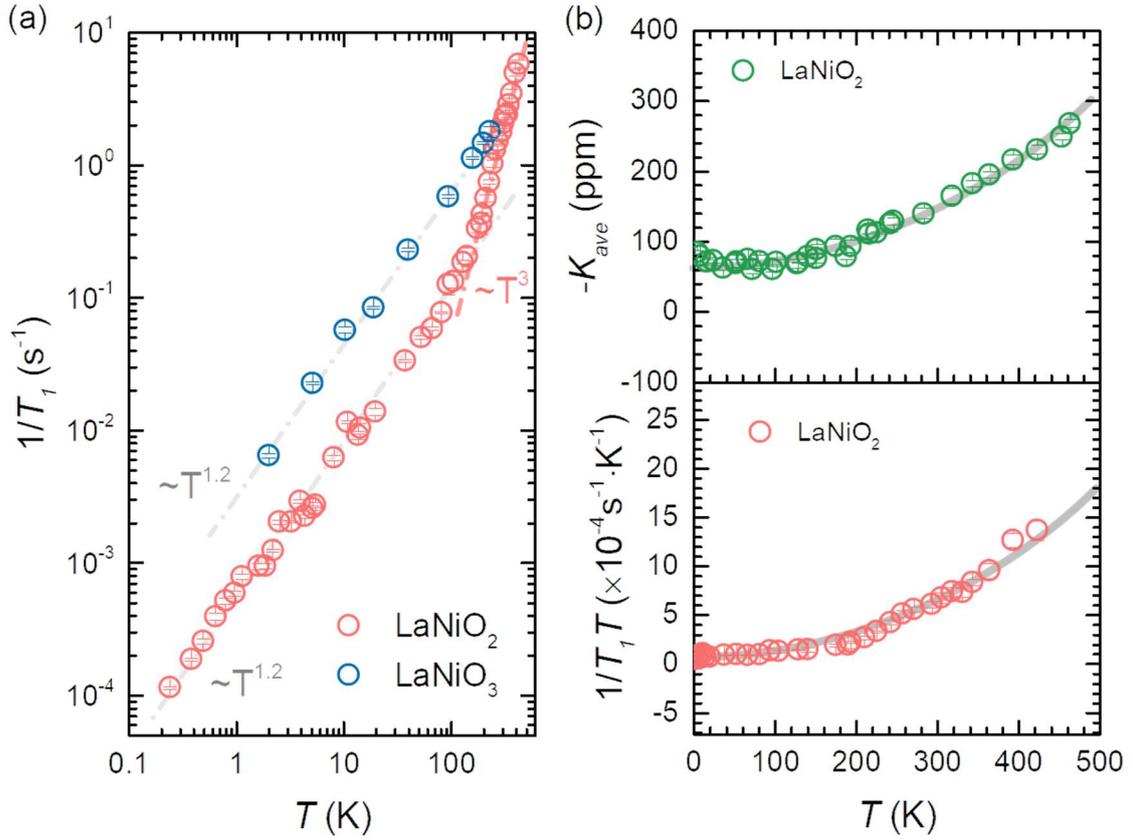

FIG. 3. Temperature-dependent Knight shift and nuclear spin-lattice relaxation in $LaNiO_2$. (a) The temperature-dependent $1/T_1$ of the $LaNiO_3$ precursor and $LaNiO_2$. (b) Upper panel: the temperature-dependent Knight shift extracted from Fig. 2(a); lower panel: the temperature-dependent $1/T_1T$ in $LaNiO_2$. A similar pseudogap-like behavior has been observed in both the Knight shift and $1/T_1T$. For a detailed discussion, see the main text. The error bars for both the Knight shift and $1/T_1$ are determined by the fitting error bar.

How to understand the similar pseudogap-like behavior in both temperature-dependent Knight shift and $1/T_1T$? In cuprates, a similar pseudogap behavior has been first revealed by $^{89}$Y NMR in YBa$_2$CuO$_{6+x}$ [37]. Later on, $^{63}$Cu and $^{17}$O NMR further confirmed such a pseudogap behavior in underdoped YBa$_2$CuO$_{6+x}$ [43-46]. A widely accepted phenomenological antiferromagnetic-Fermi-liquid theory proposed by Millis, Monien, and Pines has been used to quantitatively account for the NMR results [47]. In this model, the Korringa law has been modified, and $1/T_1T$ is proportional to $K_s$ instead of $K_s^2$. In fact, the modified Korringa relation has already been suggested by the marginal Fermi-liquid theory proposed by Varma [48]. Experimentally, such a modified Korringa relation has been successfully verified in underdoped YBa$_2$CuO$_{6+x}$ [49,50]. Then, a natural question for the present study is whether a similar modified Korringa relation also works in infinite-layer LaNiO$_2$ or not. As shown in Fig. 4(a), a linear behavior in the $1/T_1T$ vs $K_{ave}$ plot is roughly confirmed within the error bar, which seems to support a modified Korringa relation as that in underdoped cuprates. However, as shown in Fig. 4(b), the conventional Korringa law could also explain the present data very well. The only clear difference between these two scaling methods is the extrapolated orbital shift, which is about -71 ppm for modified Korringa scaling and -5 ppm for conventional Korringa scaling, respectively. At the present stage, this leaves an open issue for Korringa relation and is left for future work. Furthermore, it should be recalled that the LaNiO$_2$ powder sample possesses a quite large resistivity at room temperature (1-2 $\Omega \cdot cm$) and exhibits an insulating behavior in temperature-dependent resistivity instead of a metallic behavior as that in optimized thin film [10, 11] (see Sec. S3 in Supplemental Material [23]), which is quite consistent with a recent report on $NdNiO_2$ powder samples [20]. In addition, the $3d^9$ electronic configuration in $Ni^+$ has a higher chemical potential than that in $Cu^{2+}$, and, hence, $LaNiO_2$ has a larger charge transfer gap than that in $La_2CuO_4$ [12]. In fact, $LaNiO_2$ is theoretically predicted to be a Hubbard-type Mott insulator, not a charge transfer insulator, which was confirmed by the recent x-ray asorption sepectroscopy measurement [19]. Therefore, $LaNiO_2$ has stronger correlations than $La_2CuO_4$ due to a much larger Hubbard interaction than the charge transfer gap. These facts suggest that the interpretation of spin susceptibility should be beyond the density of states in the Fermi-liquid scenario at least in the $LaNiO_2$ powder sample. Following this idea, we could further discuss the implication of the observed pseudogap-like behavior in $LaNiO_2$.

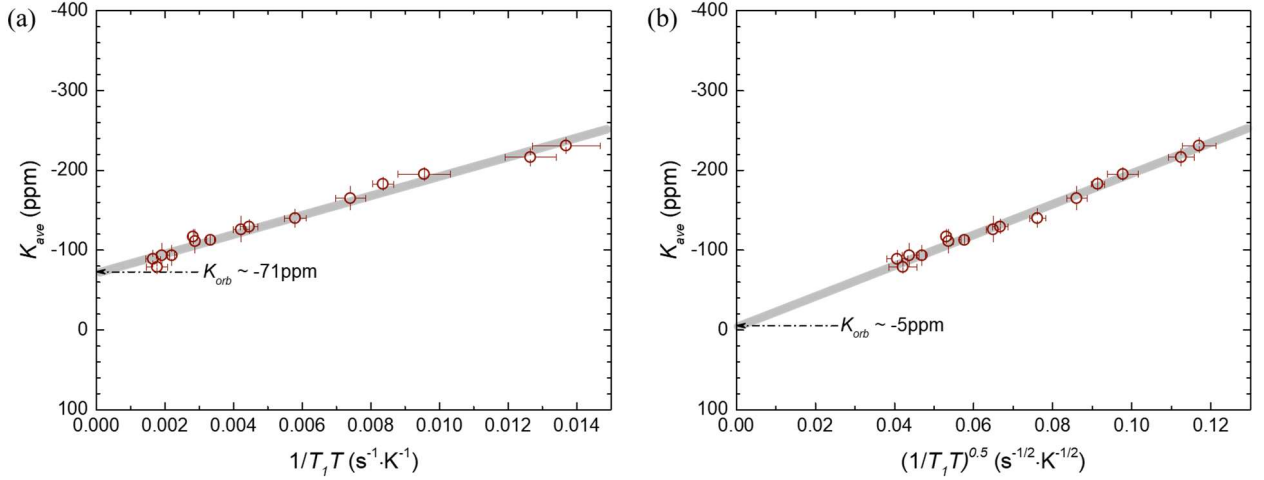

FIG. 4. Comparison of different scalings between the Knight shift and $1/T_1T$. Based on the data above 150 K in Fig. 3(b), different scalings between the Knight shift and $1/T_1T$ have been checked. (a) Modified Korringa relation between the Knight shift and $1/T_1T$; (b) conventional Korringa relation between the Knight shift and $1/T_1T$. The gray bold line indicates a linear relationship between the Knight shift and $1/T_1T$ in different scaling method. For detailed discussion, see the main text.

First, antiferromagnetic correlations should be important and play a key role in infinite-layer nickelates. Although long-range antiferromagnetic order and QCP behavior are absent in $LaNiO_2$, it does not mean that antiferromagnetic correlation is not important. In fact, a pseudogap-like behavior in spin susceptibility can be evidence for strong antiferromagnetic correlation in the two-dimensional

(2D) Heisenberg model. Based on the exact quantum Monte Carlo calculation of the 2D Heisenberg model on the square lattice, a crossover from Curie-Weiss-like behavior to pseudogap-like behavior would appear at the temperature with the value of about magnetic exchange interaction ($J$) due to the formation of short-ranged antiferromagnetic correlations [51]. In fact, a similar discussion based on the $J_1$-$J_2$ model has been also used to understand the $T$-linear behavior in the uniform spin susceptibility of iron-based superconductors [52]. When $J$ is quite large, a Curie-Weiss-like behavior would appear only at relatively high temperatures beyond room temperature. That is the reason why a Curie-Weiss-like behavior is always absent during measurement below room temperature in iron-based superconductors [53]. Only when $J$ is reduced to a smaller value could a Curie-Weiss-like behavior be observed below room temperature. This is the case for heavily hole-doped iron-based superconductor $A$Fe$_2$As$_2$ ($A$ = K, Rb, and Cs) [54,55]. In our case, the crossover temperature is higher than 460 K at least, supporting a considerable value of $J$. This would be very important for theories with Cooper pairing mediated by antiferromagnetic coupling. Second, besides the similarities with cuprates, the pseudogap-like behavior in LaNiO$_2$ also has some differences from that in cuprates. The temperature-dependent curvature of the pseudogap-like behavior in LaNiO$_2$ is actually more like that in FeSe-based superconductors, such as K$_x$Fe$_2$Se$_2$ [56] and Li$_x$(C$_2$H$_8$N$_2$)$_y$Fe$_{2-z}$Se$_2$ [57]. Although how to quantitatively explain the temperature dependence of $1/T_1T$ and the Knight shift in LaNiO$_2$ is still an open issue at the present stage, this similarity with FeSe-based superconductors might suggest a possible Hund's metal physics in infinite-layer nickelates which has been recently proposed by theory [16,18]. Especially, there are La$^{3+}$ 5$d$ electron pockets in LaNiO$_2$ [5,19]; however, their contribution in NMR measurements remains elusive. Finally, whether a similar modified Korringa relation as cuprate superconductors also exists for infinite-layer nickelates needs further verification.

In summary, by conducting a $^{139}$La NMR experiment, the magnetic ground state and intrinsic spin susceptibility have been explored in infinite-layer LaNiO$_2$. First, a paramagnetic state far from antiferromagnetic QCP has been confirmed, which is quite consistent with previous neutron scattering experiments. Second, in contrast to the previous report on bulk susceptibility, a pseudogap-like behavior instead of a Curie-Weiss-like behavior was observed. Moreover, the possible scaling behavior between the Knight shift and $1/T_1T$ has also been discussed. Finally, the present work suggests a considerable exchange interaction in infinite-layer LaNiO$_2$, which would set a strong constraint on the existing theories.

*Notes added.*–Recently, we realized that a $^1$H NMR work on doped NdNiO$_2$ polycrystalline sample has been posted [58]. Antiferromagnetism and Curie-Weiss behavior have been claimed in this work, which are quite different from the present results in LaNiO$_2$.

**Acknowledgments**

We thank the valuable discussion with L. Shu, Y. L. Wang, Z. Y. Wang and J. F. He. This work is supported by the National Key R&D Program of the MOST of China (Grants No. 2017YFA0303000 and No. 2016YFA0300201), the National Natural Science Foundation of China (Grants No. 11888101 and No. 11522434), the Strategic Priority Research Program of Chinese Academy of Sciences (Grant No. XDB25000000), and the Anhui Initiative in Quantum Information Technologies (Grant No. AHY160000).